\documentclass[10pt,letterpaper]{article}

\usepackage{opex3}
\usepackage{amsmath}
\usepackage[verbose]{cite}

\allowdisplaybreaks[1]
\begin{document}

%%%%%%%%%%%%%%%%%% title page information %%%%%%%%%%%%%%%%%%
\title{Stimulated Brillouin scattering and Brillouin-coupled four-wave-mixing in a silica microbottle resonator}

\author{Motoki Asano,$^1$ Yuki Takeuchi,$^1$ Sahin Kaya Ozdemir,$^{1,2,3}$ Rikizo Ikuta,$^1$ Lan Yang,$^2$ Nobuyuki Imoto,$^1$ and Takashi Yamamoto$^{1,4}$}

\address{$^1$ Department of Materials Engineering Science, Graduate School of Engineering Science, Osaka University, Toyonaka, Osaka 560-8531, Japan\\
$^2$Department of Electrical and Systems Engineering, Washington University, St. Louis, MO 63130, USA}

\email{$^3$ozdemir@wustl.edu}
\email{$^4$yamamoto@mp.es.osaka-u.ac.jp} %% email address is required


\begin{abstract}
We report the first observation of stimulated Brillouin scattering (SBS) with Brillouin lasing, and Brillouin-coupled four-wave-mixing (FWM) in an ultra-high-Q silica microbottle resonator. The Brillouin lasing was observed at the frequency of  $\Omega_B=2\pi\times10.4$ GHz with a threshold power of $0.45$ mW. Coupling between Brillouin and FWM was observed in both backward and forward scattering directions with separations of  $2\Omega_B$. At a pump power of $10$ mW, FWM spacing reached to 7th and 9th order anti-Stokes and Stokes, respectively.
\end{abstract}

 % REPLACE WITH CORRECT OCIS CODES FOR YOUR ARTICLE, MINIMUM OF TWO; Avoid using the OCIS codes for gGeneralh or gGeneral scienceh whenever possible.

%%%%%%%%%%%%%%%%%%%%%%% References %%%%%%%%%%%%%%%%%%%%%%%%%
%\begin{thebibliography}{99}

%%%%%%%%%%%%%%%%%%%%%%%%%%  body  %%%%%%%%%%%%%%%%%%%%%%%%%%
\section{Introduction}

Stimulated Brillouin scattering (SBS) is a nonlinear optical phenomenon that takes place due to electrostrictive interaction between light and acoustic waves (dictated by the intrinsic material nonlinearities), as well as due to the combined effect of the electrostrictive forces and radiation pressure at the boundaries (in particular in systems with nanoscale light confinement) \cite{rakich2012giant}. The process requires phase matching, which involves energy and momentum conservation \cite{agrawal2013nonlinear,kobyakov2010stimulated}. In SBS, an incident pump light with frequency $\omega_p$ is scattered from acoustic phonons of frequency $\Omega_B$ resulting in a Stokes light with a lower frequency of $\omega_S=\omega_p-\Omega_B$ (i.e., energy conservation) and amplification of the initial acoustic wave. Since the scattering can take place in any direction with respect to the pump direction, momentum conservation rule ${\bf k}_P={\bf k}_S+{\bf k}_B$ where ${\bf k}_{i=P,S,B}$ are the wavevectors of the pump light, Stokes light and the acoustic phonons, respectively, determines, together with the dispersion characteristics of the material, the frequency of the acoustic phonons. Although scattering can be in any direction in bulk systems, in light guiding structures such as optical fibers, waveguides or resonators, the pump and Stokes fields (i.e., scattering) are either co-propagating (i.e., forward scattering) or counter-propagating (i.e., backward scattering). In silica optical fibers, if the pump (at 1550 nm band) and Stokes fields are counter-propagating the acoustic frequency is 11 GHz, and if the pump and Stokes are co-propagating, it is in the range of MHz to a few GHz. SBS has been of increasing interest due to its underlying physics and potential applications such as ultra-narrow linewidth Brillouin laser \cite{smith1991narrow,li2012characterization,li2014low,loh2015dual}, Brillouin optomechanics \cite{bahl2012observation,bahl2013brillouin}, microwave photonics \cite{li2013microwave,li2014electro}, slow and fast light generation \cite{okawachi2005tunable,song2006highly}, and nonreciprocal light propagation \cite{dong2015brillouin,kim2015non}. 

Four-wave-mixing (FWM) is yet another nonlinear process that has found widespread use in many areas ranging from optical frequency conversion \cite{absil2000wavelength,carmon2007visible,turner2008ultra} and quantum light generation \cite{clemmen2009continuous,engin2013photon,azzini2012ultra,wakabayashi2015time,dutt2015chip} to quantum nondemolition measurements \cite{xiao2008quantum} and frequency combs \cite{del2007optical,savchenkov2008tunable,li2012low}. Four-wave-mixing requires the phase-matching condition  $\omega_1+\omega_2=\omega_3+\omega_4$ and ${\bf k}_1+{\bf k}_2={\bf k}_3+{\bf k}_4$ for non-degenerate process, and $2\omega_1=\omega_3+\omega_4$ and $2{\bf k}_1={\bf k}_3+{\bf k}_4$  for degenerate process.  If the phase matching conditions for SBS and FWM are simultaneously satisfied, they become coupled to each other, a process referred to as SBS-coupled FWM or SBS-assisted FWM \cite{braje2009brillouin,buttner2014phase}.

Among many optical structures and devices, microresonators, in particular, whispering-gallery-mode (WGM) microresonators, have emerged as convenient optical devices to study nonlinear processes and lasing at low input powers. SBS have been demonstrated in microspheres \cite{tomes2009photonic,guo2015low}, silica wedge resonators \cite{lee2012chemically}, silica microcapillaries \cite{bahl2013brillouin}, silica microbubble resonator \cite{lu2016stimulated}, $\mathrm{BaF_2}$ disk resonator \cite{lin2015opto} and in crystalline (e.g., $\mathrm{CaF_2}$ and $\mathrm{BaF_2}$) resonators \cite{grudinin2009brillouin,lin2014cascaded}. In \cite{lu2016stimulated}, SBS and FWM were reported as two separate (not simultaneous or in a way to enhance/suppress each other) nonlinear processes taking place in a hollow-core microbottle. In \cite{lin2015opto}, the anti-Stokes components in the forward and backward directions hint the possibility of a coupled SBS and FWM process (via FWM between the pump and the Stokes generated by SBS or FWM of Stokes); however, the strong presence of the pump in both the forward and backward (due to strong Rayleigh backscattering) directions does not convincingly exclude the possibility of the observed anti-Stokes components coming solely from FWM. In these structures, Brillouin interaction is significantly enhanced because they support both optical and acoustic resonances. Similarly, the FWM has been investigated in WGM resonators \cite{absil2000wavelength,kippenberg2004kerr,savchenkov2004low,agha2007four}. The presence of doubly or triply resonant optical modes in microresonators enhance FWM interactions leading to a more efficient process.

Recently solid-core microbottle resonators have attracted increasing interest due to their easy fabrication, high optical quality factors ($Q\leq 10^8$) and 3D optical confinement \cite{sumetsky2004whispering,pollinger2009ultrahigh}. Experiments of cavity-QED \cite{volz2014nonlinear}, add-drop filter \cite{murugan2010optical}, slow light generation \cite{sumetsky2013delay} and nonlinear optical Kerr switch \cite{pollinger2010all} have shown that microbottles are promising for applications and fundamental studies. In contrast to other WGM resonators, the free-spectral-range (FSR) of microbottle resonators depend on not only the radius of the structure but also its axial length \cite{louyer2005tunable}, making it possible to fabricate resonators with a preferred density of resonances such that there are resonances at the frequencies of interest (i.e., resonances at the pump and Stokes frequencies for cavity-enhanced SBS). 

In this paper, we report the first observation of SBS, low-threshold Brillouin lasing and SBS-coupled FWM in a silica microbottle resonator, fabricated from a single mode optical fiber using heat-and-pull method.
 
\section{Principle and experimental setup}
In cavity-enhanced SBS, the resonator should support both optical resonances at the frequency of the pump and Stokes light (i.e., the FSR of the resonator corresponds to the Brillouin frequency $\Omega_B$), and the acoustic wave. Figure 1(a) depicts an illustration of SBS process. The 1st Stokes signal S1 red-shifted from the pump by $\Omega_B$ is the result of backward scattering of the pump from the acoustic wave. At elevated pump powers, the 1st Stokes signal is enhanced such that it acts as a pump that generates the 2nd Stokes signal via backward scattering. Then the 2nd Stokes, which propagates in the same direction as the initial pump, creates the 3rd Stokes which co-propagates with the 1st Stokes but counter-propagates with the initial pump and 2nd Stokes.  If the power is further increased, high-order Stokes fields at $\omega_{nS}=\omega_p-n\Omega_B$ are generated. Since the odd (even) order Stokes counter-propagates (co-propagates) with the initial pump field, a train of Stokes signals with a frequency spacing of $2\Omega_{B}$ emerges in the forward (backward) directions, forming a Brillouin frequency comb. 

In cavity-enhanced FWM, the resonator should support resonances at the pump, Stokes (red-shifted from the pump; idler) and anti-Stokes (blue-shifted from the pump; signal) frequencies. Figure 1(b) presents an illustration of degenerate and non-degenerate FWM with the idler, pump and signal separated from each by one FSR ($\Omega_{FSR}$). Thus, a cascaded FWM process can generate high-order Stokes and anti-Stokes fields at $\omega_{nS}=\omega_p-n\Omega_{FSR}$ and $\omega_{naS}=\omega_p+n\Omega_{FSR}$. This can be considered as a FWM frequency comb with a spacing of $\Omega_{FSR}$. The phase matching condition for FWM requires that all the fields involved in the process co-propagate in the resonator. Note that presence of anti-Stokes is a signature of FWM because SBS cannot generate anti-Stokes. 

Figure 1(c) illustrates the formation of SBS-coupled FWM and frequency comb. In order for SBS and FWM to assist and enhance each other, phase matching for both processes should be satisfied simultaneously. For this to take place in a resonator, the FSR of the resonator should be equal to the Brillouin frequency $\Omega_B$. Although independently (not coupled) excited SBS (or FWM) can generate Stokes (or Stokes and anti-Stokes) signals separated from each other by the frequency $\Omega_B=\Omega_{FSR}$, SBS-coupled FWM generates signals separated from each other by $2\Omega_{B}$ as dictated by the requirement of simultaneous phase matching [Fig. 1(c)]. Let us assume that high-order SBS created the 1st, 2nd and the 3rd Stokes at $\omega_{S1},\omega_{S2}$ and $\omega_{S3}$ separated by $\Omega_B$. The 1st and the 3rd Stokes are in the same direction, counter-propagating with the initial pump, whereas the 2nd Stokes co-propagates with the pump. In this case 2nd Stokes and the pump can undergo non-degenerate FWM creating the 2nd anti-Stokes at  $\omega_{aS2}=\omega_P+2\Omega_{B}$ and the 4th Stokes at $\omega_{S4}=\omega_{S2}-2\Omega_{B}$, which co-propagate with the pump and the 2nd Stokes. Similarly the 1st and the 3rd Stokes can undergo FWM creating the 1st anti-Stokes at $\omega_{aS1}=\omega_{S1}+2\Omega_{B}$ and the 5th Stokes at $\omega_{S5}=\omega_{S3}-2\Omega_{B}$ which co-propagate with the 1st and 3rd Stokes but counter-propagates the initial pump. Note that any of the generated Stokes or anti-Stokes can also undergo degenerate FWM creating new Stokes and anti-Stokes. If the SBS-coupled FWM continues, with the newly generated signals acting as seeds, more Stokes and anti-Stokes signals with new frequencies will be generated. As a result, if the out-coupled light from the waveguide-coupled resonator is observed, one will see a comb with spacing $2\Omega_B$ in both the forward (i.e., the same direction with the initial pump) and the backward (i.e., the opposite direction with the initial pump) directions. While the former includes Stokes and anti-Stokes of even order, the latter includes only the odd order. We note that a frequency comb via SBS-only process consists of Stokes signals (frequencies smaller than spacing of $2\Omega_B$. A comb via FWM-only process, on the other hand, consists of Stokes and anti-Stokes signals (frequencies smaller and larger than the pump frequency) with a spacing of $\Omega_B$. Combining these two process as SBS-coupled FWM prepares a frequency comb consisting of Stokes and anti-Stokes (i.e., similar to a FWM-only process) with a frequency spacing of  $2\Omega_B$ (i.e., similar to a SBS-only process). Here, we ignore the spectra from SBS-uncoupled FWM, which generates signals and idlers with a separation of $\Omega_{FSR}$ because Brillouin gain is about one hundred times larger than FWM gain \cite{agrawal2013nonlinear}.   

\begin{figure}[htbp]
\centering\includegraphics[width=0.9\linewidth, angle=0]{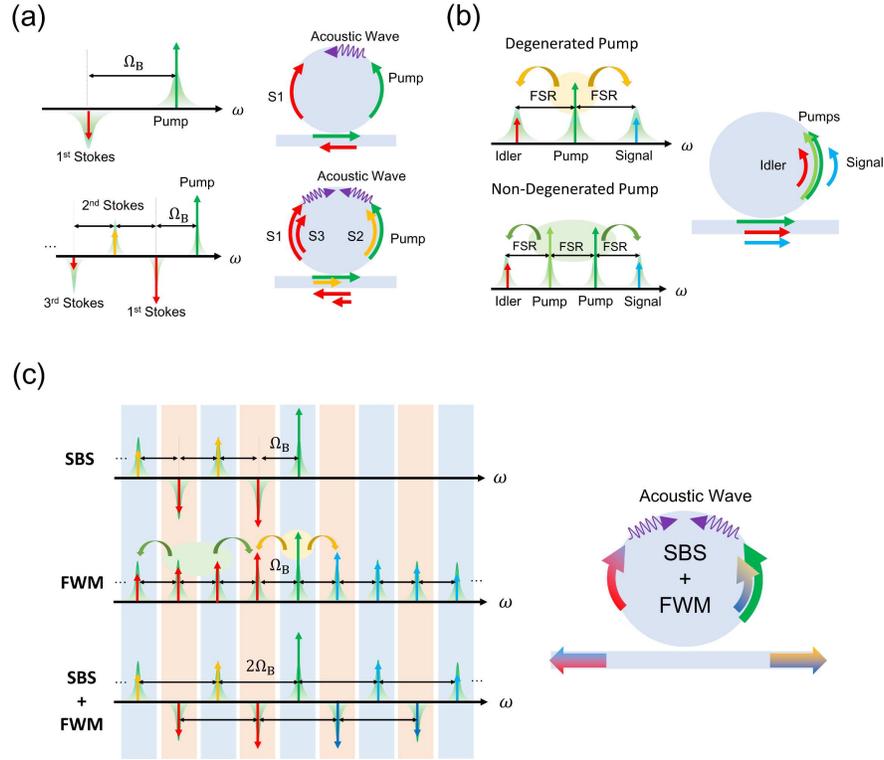}
\caption{Illustration of the mechanisms of SBS, FWM, and SBS-coupled FWM in a waveguide-coupled resonator. Downward (upward) arrows denote signals generated in the opposite (same) direction with the pump. (a) SBS-only process generates only Stokes signals separated from each other by the Brillouin frequency $\Omega_B$=$\Omega_{FSR}$. The even-order Stokes co-propagates with the pump whereas odd-order Stokes counter-propagates. (b) FWM-only process generates both Stokes and anti-Stokes with a spacing of free-spectral-range $\Omega_{FSR}$ of the resonator. All signals co-propagate with the pump. (c) SBS-coupled FWM requires simultaneous phase matching for both SBS and FWM. Both Stokes and anti-Stokes signals are generated with a spacing of $2\Omega_B$. In (c) the signals in the columns with the same color undergo degenerate or non-degenerate FWM.}
\end{figure}

We used a silica microbottle resonator, which was fabricated from a single mode optical fiber (125 $\mathrm{\mu}$m clad and 10 $\mathrm{\mu}$m core) using the heat-and-pull method, to study SBS, Brillouin lasing and SBS-assisted FWM. The maximum diameter of the bottle resonator, the diameter of the necks and the separation between the necks were approximately 125 $\mathrm{\mu}$m, 80 $\mathrm{\mu}$m and 8 mm, respectively. Figure 2(a) shows the optical microscope image of the microbottle resonator used in the experiment. Using a theoretical model \cite{louyer2005tunable}, we estimated the azimuthal FSR as $\sim$0.5 THz, and the axial FSR as $\sim$10 GHz that is roughly equal to $\Omega_B$. From the transmission spectra measured with a weak probe when the taper-resonator system was set in the deep-undercoupling regime, we estimated the quality factor of the resonances as $2.2\times10^8$. Figure 2(b) depicts an illustration of the experimental setup. Light from an external cavity diode laser (ECDL) in the 1550 nm band with a linewidth of 300 kHz was used to probe the resonance structure of the microbottle and to excite the SBS and FWM in the resonator. The wavelength of the ECDL was linearly scanned by applying a triangular signal, generated by an arbitrary function generator (AFG) to the piezoelectric transducer inside the laser package. Before the light from the ECDL was launched into the tapered fiber, it was first amplified using an erbium-doped fiber amplifier (EDFA), and then sent through a tunable attenuator (Att) to have a tunable pump power. After the attenuator, the polarization of this light was adjusted by a fiber polarization controller (PC) for maximal coupling into the WGM. Using a 10:90 fiber-based beamsplitter (BS), a portion of the light was directed to a power meter. The rest (larger portion) was coupled into a tapered fiber that was used to couple light into and out of the WGMs of the microbottle. The forward signal (transmission) was divided into two optical paths, one of which was monitored by a photodiode (PD) connected to a digital sampling oscilloscope (DSO) and the other was connected to a single channel optical spectrum analyzer (OSA). In order to monitor the backward signal (reflection), a circulator was placed between the 10:90 beamsplitter and the tapered fiber so that the back-scattered light could be channeled to the OSA and to a PD connected to the DSO.  We placed a fiber-based switch to actively route either the forward or the backward signal to the single channel OSA. We fixed the laser-resonator detuning using thermal locking method. 
\begin{figure}[htbp]
\centering\includegraphics[width=0.9\linewidth]{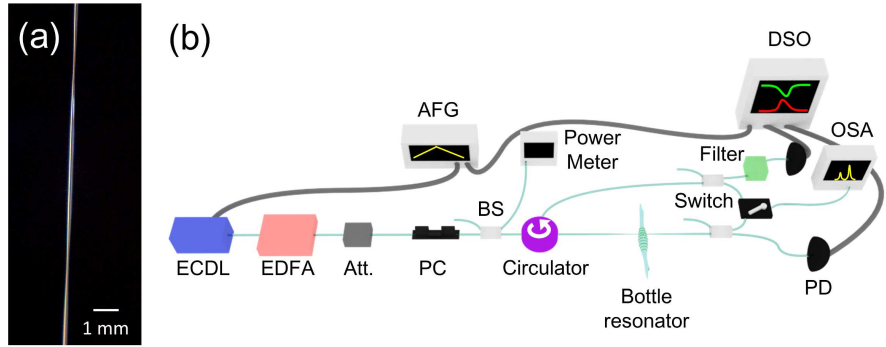}
\caption{(a) Optical microscope image of the microbottle resonator used in the experiment.(b) Experimental setup. ECDL: External cavity diode laser; AFG: arbitrary function generator; EDFA: Erbium-doped fiber amplifier; Att: Attenuator; PC: Polarization controller; BS: beamsplitter; PD: photodiode; DSO: digital sampling oscilloscope; OSA: Optical spectrum analyzer. }
\end{figure}
\section{Results and discussion}
Figures 3(a) and 3(b) show the spectra obtained in the backward and the forward directions at a pump power of 1.6 mW.  As expected the 1st Stokes, that was red-shifted from the pump by the Brillouin frequency of $\Omega_B=2\pi\times$10.9 GHz, appeared in the backward spectra. As we increased the pump power and adjusted the detuning, cascaded SBS took place at 5.1 mW [Figs. 3(c)-3(d)], generating the 1st and 3rd Stokes in the backward and the 2nd Stokes in the forward spectra. 
\begin{figure}[htbp]
\centering\includegraphics[width=0.9\linewidth]{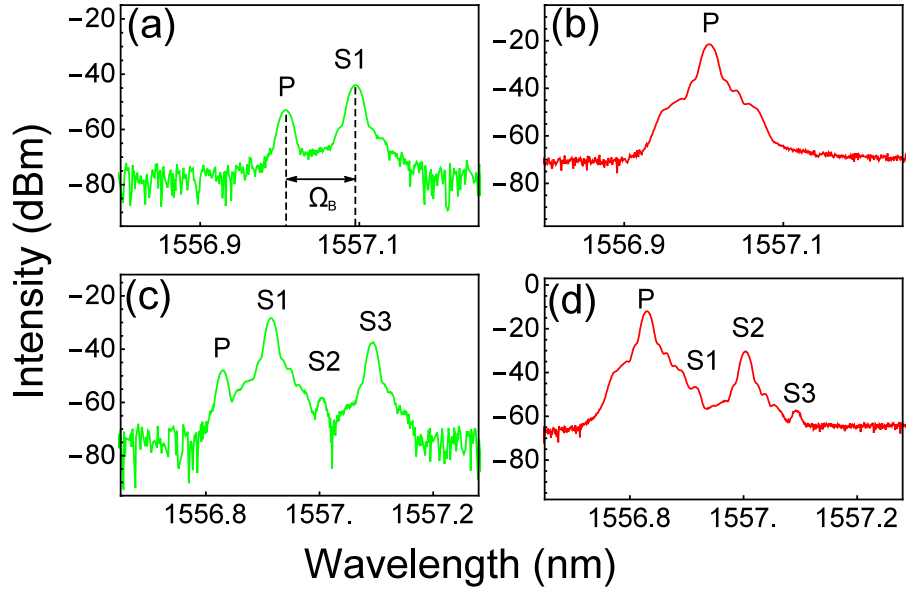}
\caption{The backward (a)-(c) and the forward (b)-(d) SBS spectra monitored with an OSA. At a pump power of 1.6 mW, only the 1st Stokes (S1) red-shifted by $\Omega_B$ from the pump (P) was observed in the backward direction (a). No Stokes was observed in the forward direction (b). At a pump power of 5.1 mW (c)-(d), cascaded SBS generated up to 3rd Stokes (S1, S2 and S3). Presence of S2 in the backward (c) and of S1 and S3 in the forward (d) directions implies Rayleigh scattering which couples backward light to the forward direction and vice versa. }
\end{figure}

As we discussed in the preceding paragraphs [Fig. 1(a)], the even-order Stokes  generated via SBS should be in the forward direction (i.e., co-propagating with the pump) and the odd-order Stokes in the backward direction (i.e., counter-propagating with the pump). In the experimentally-obtained spectra of cascaded SBS given in Figs. 3(c)-3(d); however, we see that there is a weak (compared to 1st Stokes) 2nd Stokes signal in the backward spectra. Similarly we see the 1st and 3rd Stokes in the forward spectra, which should not have odd-order Stokes. These suggest the presence of Rayleigh scattering that may be due to the presence of scattering centers in the mode volume. In such cases, either the strong pump field input in the forward direction scatters into the backward direction and generates Stokes both in the backward and forward directions or the 1st and 3rd Stokes generated in the backward direction undergo Rayleigh scattering and couple to the forward direction. In the case of very strong Rayleigh scattering, the  odd- and even-order Stokes in both the forward and backward directions can@become comparable to each other, leading to an output with a frequency spacing of $\Omega_B$ in both the forward and backward directions, instead of the frequency spacing of $2\Omega_B$ expected when there is no Rayleigh scattering [Fig. 1(c)]. 

In order to observe the Brillouin lasing behavior for the 1st Stokes signal, the backward and the forward signals were monitored by the DSO by scanning the laser frequency for $\sim$3 GHz around a resonant mode. We inserted an optical bandpass filter, with a bandwidth of 60 GHz around the pump frequency, into the path of the backward scattering before the PD to remove signals generated by other nonlinear processes such as Raman and FWM. Figures 4(a) and 4(b) depict the experimentally-obtained transmission spectra (forward scattering) and reflection spectra (backward scattering) for the pump powers of 0.8 mW and 3.2 mW, respectively. In the forward direction, we have a thermally-broadened transmission spectra due to ultra-high Q resonance and high pump power. We observed a signal peak in the backward direction at the edge of the transmission spectrum. The intensity of this signal increased as we increased the pump power. We evaluated the dependence of this signal on pump power by monitoring it with the PD and the OSA simultaneously. From the OSA data we found that this signal was red-shifted from the pump by 10.5 GHz which equals to the Brillouin frequency of $\Omega_B$, suggesting that the observed signal was the 1st Stokes [Fig. 4(c)]. As seen in Fig. 4(c) pump power versus peak power of the 1st Stokes exhibited a threshold behavior. From a linear fit to the experimental data, we estimated the threshold of Brillouin lasing as 0.45 mW.
\begin{figure}[htbp]
\centering\includegraphics[width=0.9\linewidth]{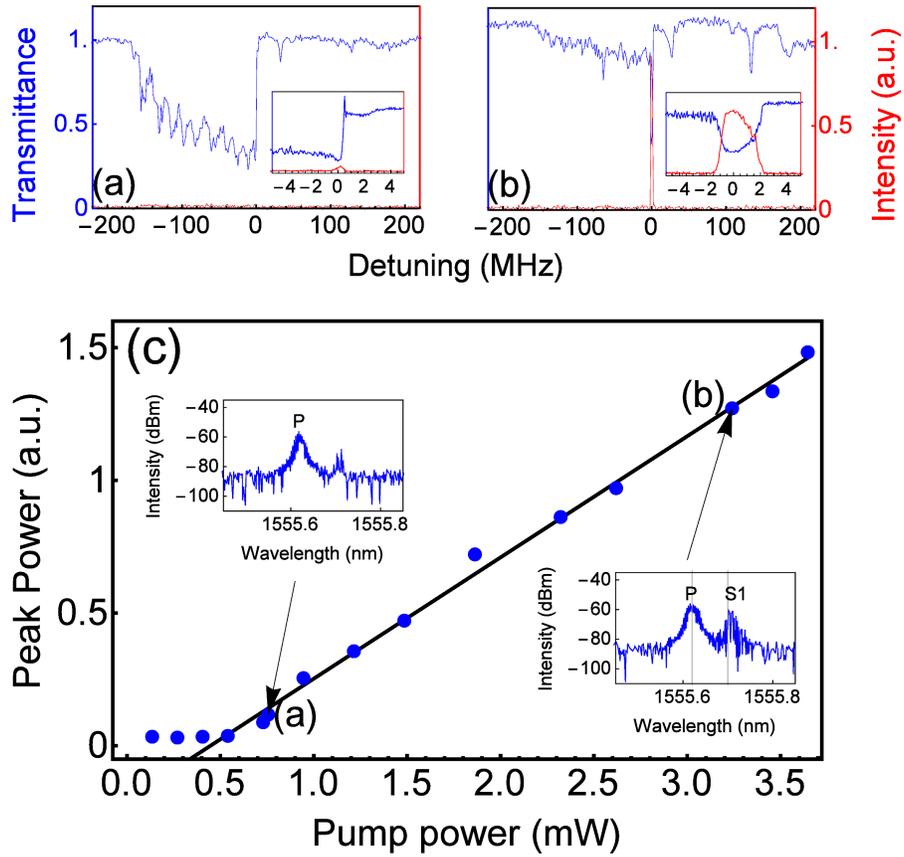}
\caption{Brillouin lasing in a microbottle. (a),(b) Experimentally-obtained spectra for a pump power of 0.8 mW (a) and 3.2 mW (b), respectively. Blue and red spectra were obtained in the forward and the backward directions. The insets show the enlarged spectra. (c) The pump power vs the peak power of the backscattered signal. The inset shows the OSA spectra around and after the lasing threshold respectively. The blue dots are the experimentally-obtained data and the solid black line denotes the best linear fit to the experimental data.}
\end{figure}

By finely adjusting the frequency of the pump, we obtained with a pump power of 6.4 mW the 1st anti-Stokes in addition to the 1st, 2nd and 3rd Stokes in the backward spectra [Fig. 5(a)], implying the presence of both FWM and cascaded SBS. The forward spectra included only the 1st and 2nd Stokes. The presence of weak 1st Stokes and weak 2nd Stokes in the forward and backward spectra, respectively, suggests the presence of Rayleigh scattering. The process here can be explained as follows. As soon as the 1st Stokes is generated by the pump in the backward direction, it experiences degenerate FWM ($2\omega_{S1}=\omega_{aS1}+\omega_{S3}$) creating the 3rd Stokes and the 1st anti-Stokes with $\omega_{S1}-\omega_{aS1}=\omega_{S3}-\omega_{S1}=2\pi\times$21.3GHz which is twice the Brillouin frequency of $\Omega_B=2\pi\times$10.7 GHz. Figure 5(b) shows the spectra obtained with a pump power of 10.0 mW. The presence of high order Stokes (up to the 11th) and anti-Stokes (up to the 5th) in the backward spectrum implies that cascaded-SBS and -FWM took place.  In the forward spectrum, anti-Stokes up to the 7th order and Stokes up to the 9th order are clearly seen. The presence of even-order signals in the backward spectrum and odd-order signals in the forward spectrum imply not only the presence of Rayleigh scattering but also the SBS-uncoupled FWM process. In the SBS-uncoupled FWM process, the strong pump light can generate the signal and idler in the forward direction with the FSR $\Omega_B$. The odd- order signals in the backward spectrum have a frequency separation of $\Omega_2=2\Omega_B=2\pi\times$21.3 GHz. This together with the presence of anti-Stokes suggests an SBS-coupled FWM. In the forward direction, there is a significant effect of Rayleigh scattering which has led a frequency separation of  $\Omega_B=2\pi\times$10.7 GHz up to 5th anti-Stokes and 7th Stokes.
\begin{figure}[htbp]
\centering\includegraphics[width=0.9\linewidth]{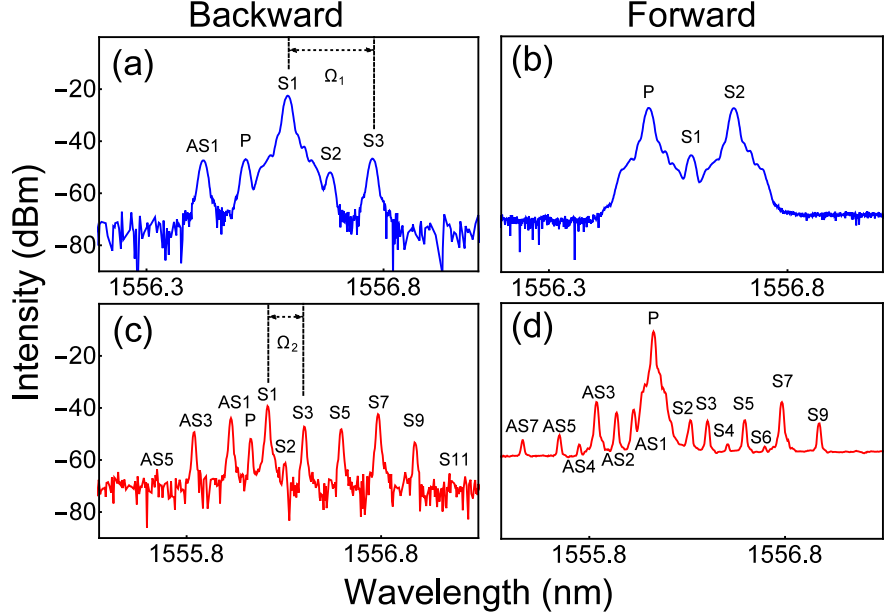}
\caption{SBS-coupled FWM observed in the backward (a),(c) and forward (b),(d) spectra obtained by an OSA. The spectra given in (a),(b)  and (c),(d) were obtained at a pump power of  6.4 mW and 10.0 mW respectively.}
\end{figure}

\section{Conclusion}
In conclusion, we reported the first observation of SBS, cascaded SBS, Brillouin laser and SBS-coupled FWM in a microbottle resonator. A challenge in the demonstration of Brillouin scattering in resonators is to precisely match the FSR of the resonator to the Brillouin frequency shift. This requires an FSR of around 11 GHz to observe backward Brillouin scattering in silica resonators. In order to overcome this challenge, resonators with millimeter-scale radius are fabricated because larger radius results in shorter FSR and a denser spectra which makes it easier to resonantly collect Brillouin scattering. In bottle resonators, on the other hand, this can be achieved by engineering the FSR of axial modes via the axial length of the bottle (i.e., the longer the axial length, the shorter the FSR and denser the spectra) while keeping the radius of the resonator at micrometer scale. This property of microbottle resonators will be very useful in studying nonlinear phenomena which requires doubly- or triply-resonant interactions. In this study, we observed SBS-coupled FWM in a resonator of radius 125 $\mathrm{\mu m}$ at pump powers as low as 6 mW, which is less than 1/10-th of the pump power (80 mW) used in BaF2 disk resonator with a diameter of 12 mm \cite{lin2015opto}, in which the possibility of SBS-coupled FWM was claimed. The high-Q, the ease of fabrication, the ability of tuning the FSR and the low power threshold for observing nonlinear effects such as FWM and Brillouin scattering and lasing make microbottle resonators suitable for many applications including sensors. For example, microbottle resonators with longer axial length and hence shorter FSR may have such a dense spectra that temperature- and strain-dependence of Brillouin frequency shift can be monitored to build optical sensors. Our results show that microbottle resonators provide a new and versatile platform to build optical sensors and optical frequency combs and Brillouin optomechanics.

\section*{Acknowledgments}
This work was supported by MEXT/JSPS KAKENHI Grant Number 16H01054, 16H02214, 15H03704, 15KK0164. We thank to the supports from Program for Leading Graduate Schools: "Interactive Materials Science Cadet Program".

\end{document}